\documentclass[epsfig,12pt]{iopart}
\usepackage{epsfig}
\usepackage{iopams}

\begin{document}

\title[A Gravitational shock wave in quadratic gravity]
{A Gravitational shock wave generated by a beam of null
matter in quadratic qravity}

\author[de Rey Neto]{E. C. de Rey Neto, J.C.N. de
Araujo, O. D. Aguiar}

\address{Instituto Nacional de Pesquisas Espaciais - Divis\~ao de
Astrof\'\i sica \\ Av. dos Astronautas 1758, S\~ao Jos\'e dos
Campos, 12227-010 SP, Brazil}

\ead{\mailto{edgard@das.inpe.br},
     \mailto{jcarlos@das.inpe.br}, and
     \mailto{odylio@das.inpe.br}}

\begin{abstract}
In the
present work we approximate an ultrarelativistic jet by a
homogeneous beam of null matter with finite width. Then, we study
the influence of this beam over the spacetime metric in the
framework of higher-derivative gravity. We find an exact shock
wave solution of the quadratic gravity field equations and compare
it with the solution to Einstein's gravity. We show that the
effect of higher-curvature gravity becomes negligible at large
distances from the beam axis. We also observe that only the Ricci-squared term 
contribute to modify the Einstein's gravity prediction. Furthermore, we note 
that this
higher-curvature term contribute to regularize the
discontinuities associated to the solution to Einstein's general
relativity. 
\end{abstract}



\maketitle

\newpage

\section{Introduction}

The relativistic and ultrarelativistic jets arise as important
components in the structure and dynamics involved in several
astrophysical scenarios. Among the different types of
astrophysical jets, the most energetic ones are potential
candidates to give rise to emission of gravitational waves. For
example, highly relativistic jets should be associated with some sources of 
gamma
ray bursts (GRBs). In the last few years, observations that indicate the 
presence
of jets in
GRBs have been reported by many authors~\cite{saripiran99}. The expansion of the
jet in the burst
leads to very high Lorentz factors which is at least of hundreds,
but higher values such as $\Gamma\sim 10^3\!-\!10^{6}$ are taken
into account in modelling the GRB sources~\cite{usov94}. In the
present paper, we consider a model in which the gamma Lorentz
factor takes the limit $\Gamma\rightarrow\infty$. This assumption
can be used to solve exactly the gravitational field equations.
Then, we can take this analytic solution to represent the
gravitational wave generated in a strong burst. To study the
impact of an ultrarelativistic jet over the spacetime metric, we
start from the extreme situation where the velocity of the
particles in the beam is assumed to be equal to the velocity of
light. The jet is then represented by a beam of null particles.
For jets which start with a small opening angle $\theta_0\leq
10^{-3}-10^{-4}$~\cite{saripiran99}, we assume that the width of
the beam remains constant during the first stage of the jet
expansion. Then, we calculate the effect of
this jet  over the spacetime metric in a flat background. We take advantage of 
the
use of exact methods to provide solutions to higher-derivative
gravity equations~\cite{buch} and compute the solution to
quadratic gravity field equations.

Quadratic gravity is an example of the higher order theories of
gravity which are generally covariant extensions of the general
relativity. These theories are constructed by the addition of
terms nonlinear in the curvature to the Einstein-Hilbert action.
The coupling parameters of the quadratic
curvature terms in the Lagrangian must be determined by
experiments. Observational 
constraints on these parameters can be traced out from sub-millimeter tests of 
the inverse square law~\cite{subexp} or by the bending of light by a gravitational field~\cite{accgr}. Are there a way to obtain an observational constraint 
on quadratic gravity coupling parameters from gravitational radiation?
The application of the traditional methods of ordinary
gravitational wave theories, which deals with the linearized version of the gravitational theories, are until the present date, unable to clarify the role played
by the quadratic curvature terms in the gravitational radiation~\cite{will1}. In a recent paper~\cite{toap} we deal with the linearized version of quadratic gravity treating the higher-curvature 
terms perturbatively.  We conclude that the effects of quadratic curvature terms 
are so small that cannot be measured by the current interferometric or mass 
resonant detectors~\footnote{The amplitude of a general linearized oscillatory quadratic gravity 
wave  at a given distance from the source differ from
the Einstein's linearized wave amplitude by a frequency dependent function. For a wave with frequency of 100 {\rm Hz} and considering an upper bound of $10^{-4} {\rm cm}^2$ for $|\beta|$~\cite{accgr}, this difference becomes~$\lesssim 10^{-21}$. This difference cannot be measured by the current gravitational wave detectors.}. 

In the present paper, we obtain a solution which can be able to explain the effect of higher-curvature
invariants on gravitational radiation by dealing with exact
solutions to gravitational field equations.
We obtain exact shock wave 
solutions. A gravitational shock wave is truly a discontinuity of the spacetime 
metric which propagates with the velocity of light in a given 
direction~\cite{baris1}. The notion of wave amplitude as a finite quantity which 
varies with the distance from the source cannot be applied here. The 
gravitational shock wave must be taken instead as a propagating singularity. 
According to~\cite{barhog1} a modification of the gravitational wave detectors 
would be necessary to observe gravitational shock waves.
We do not intend to discuss the detection of gravitational shock waves in the 
present work. The tidal field of a gravitational shock wave is an issue which we 
are dealing with and will be presented in a forthcoming paper. 

The plane of the present paper is the following.
In section~\ref{sec-2} we write the field equations for the shock
wave metric in the frameworks of quadratic gravity and Einstein's
general relativity. This is carried out by assuming
that the solution can be described by a pp-wave
metric~\cite{Kramer80}. In section~\ref{sec-3} we consider a
homogeneous and finite beam of null particles and obtain the
gravitational shock wave generated by it. We also compare
the solutions obtained in both theories. In section~\ref{conc} we
summarize the principal results obtained and make some comments concerning to 
the 
application of these results in astrophysics.

\section{The field equations for a plane gravitational shock wave}
\label{sec-2}

We start with an impulsive pp-wave metric given by the following
line element
\begin{eqnarray}
ds^2=-dudv+H(x^i,u)du^2+(dx^i)^2, \label{shockmetric}
\end{eqnarray}
where
\begin{eqnarray}
H(x^i,u)=f(x^i)\delta(u),\;\;\;\;\;\;u=t-z,\;\;\;\;\;v=t+z,
\end{eqnarray}
and $x^i$ (i=1,2) denote the Cartesian coordinates in the plane perpendicular to 
the wave propagation, which we call transverse coordinates. The 
metric~(\ref{shockmetric}) represents a gravitational shock wave propagating in 
$z$ direction. This is a
variation of the Peres's wave metric~\cite{peres} and was used by
many authors to study wave-like exact solutions to Einstein
gravitational equations~\cite{buch,lou1}.
The coefficient of $\delta$-distribution is the wave profile function. This
function must be determined by the field equations and will depend on the
characteristics of the source and on the underlying theoretical model.

The nonvanishing Christoffel symbols for the metric~(\ref{shockmetric}) are
given
by
\begin{eqnarray}
\Gamma^v_{uu}=-f\dot{\delta},\;\;\;\;\;\Gamma^i_{uu}=-\frac{1}{2}\partial_i
f\delta,\;\;\;\;\;
\Gamma^v_{ui}=\Gamma^v_{iu}=-\partial_if\delta,\label{christ}
\end{eqnarray}
where $\partial_i$ denotes partial derivatives with respect to the
coordinate $x^i$, and the dot denotes the derivative  with respect to the
$u$ coordinate.
The only nonvanishing components of the Riemann tensor, apart
from the ones obtained by symmetry properties, are given by

\begin{eqnarray}
R_{iuju}=-\frac{1}{2}\partial_i\partial_jH(x^i,u).\label{Riuju}
\end{eqnarray}
The only nonvanishing components of the Ricci tensor for the
metric~(\ref{shockmetric}) are

\begin{eqnarray}
R_{uu}=-\frac{1}{2}\nabla^2_\perp H(x^i,u),\label{Ruu}
\end{eqnarray}
where $\nabla^2_\perp$ denotes the Laplacian operator in the
transverse space $\{x^i\}$. The curvature scalar $R$ and the
elementary quadratic invariants vanish identically

\begin{eqnarray}
R=0,\;\;\;\;R_{\alpha\beta}R^{\alpha\beta}=0\;\;\;\;R_{\alpha\beta\gamma\delta
}R^{\alpha\beta\gamma\delta}=0.\label{invvanish}
\end{eqnarray}

The quadratic gravity theory in four dimensions can be derived from the action
\begin{eqnarray}
S =\frac{1}{16\pi G}{\int d^4x\sqrt{-g}\{R+\alpha
R^2+\beta R_{\mu\nu}R^{\mu\nu}+16\pi G {\cal L}_m}\}.\label{ac1}
\end{eqnarray}
Since, in four spacetime dimensions, the Gauss-Bonnet invariant
can be used to eliminate the $R_{\alpha\beta\gamma\delta
}R^{\alpha\beta\gamma\delta}$ term. In the above action, ${\cal
L}_m$ stands for the presence of matter fields. The units are such
that $\hbar=c=1$. We remark that the gravitational actions built from Lagrangian 
densities  which are arbitrary functions of $R$  are conformally equivalent to 
general relativity interacting with scalar fields. However, this 
equivalence cannot be traced for the theory derived from the action~(\ref{ac1}) 
due to the presence of the $R_{\mu\nu}R^{\mu\nu}$ invariant~\cite{gott}. 

Requiring the action $S$ to be stationary leads
to the following field equations

\begin{eqnarray}
R_{\mu\nu}-\frac{1}{2}g_{\mu\nu}R+\alpha {\rm H}_{\mu\nu}+\beta
{\rm I}_{\mu\nu}=8\pi GT_{\mu\nu},\label{fieldeq}
\end{eqnarray}
where
${\rm H}_{\mu\nu}$ and ${\rm I}_{\mu\nu}$ are the operators associated 
respectively with the $R^2$ and $R_{\mu\nu}R^{\mu\nu}$ invariants and are 
given by
\begin{equation}
{\rm H}_{\mu\nu}=-2 R_{;\mu\nu}+2g_{\mu\nu}\opensquare
R-\frac{1}{2}g_{\mu\nu}R^2+2RR_{\mu\nu},
\end{equation}
and
\begin{eqnarray}
{\rm I}_{\mu\nu}=-2R^\alpha_{\;\;\mu;\nu\alpha}+\opensquare
R_{\mu\nu}+\frac{1}{2}g_{\mu\nu}\opensquare
R+2R^\alpha_{\;\;\mu}R_{\alpha\nu}-
\frac{1}{2}g_{\mu\nu}R_{\alpha\beta}R^{\alpha\beta},
\end{eqnarray}
where $\Box$ refers to the curved space d'Alambert operator.
The theory defined by the action~(\ref{ac1}) leads to the
following non-relativistic gravitational potential~\cite{acptp}:
\begin{eqnarray}
V(r)=GM\left\{-\frac{1}{r}+\frac{4}{3}\frac{e^{-m_1r}}{r}-
\frac{1}{3}\frac{e^{-m_0r}}{r}\right\},\label{potnr}
\end{eqnarray}
where
\begin{eqnarray}
m_0^2=\frac{1}{3\alpha+\beta},\;\;\;\;\;\;\;m_1^2=-\frac{1}{\beta}.
\end{eqnarray}

This potential reflects the three modes of the linearized theory,
one with a vanishing mass which gives the Newtonian force and two
massive modes which create Yukawa-type interactions. To obtain an
acceptable Newtonian limit we restrict $m_0$ and $m_1$ to real
values, leading to the, so called, no-tachyon constraints 
\begin{eqnarray}
3\alpha+\beta\geq 0,\;\;\;\;\;\;\;-\beta\geq 0. \label{ntachy}
\end{eqnarray}

Experiments which tests the inverse square law at sub-millimeter distances indicate an upper limit to the absolute value 
of the $\alpha$ or $\beta$ parameters
of $10^{-4} {\rm cm^2}$~\cite{subexp}. A direct observational constraint on the $\beta$ parameter can be obtained from the bending of light by the Sun's gravitational field.  According to a semiclassical computation of the 
bending of light in quadratic gravity~\cite{accgr}, the higher-curvature 
coupling parameter $\beta$ must satisfy the same constraint, namely  
$|\beta|<10^{-4}{\rm cm^2}$. This  bound improves by several orders of magnitude 
the early accepted value for the upper limit on 
$|\beta|$~\cite{stelle78}.

Now, let us write the field equations for the metric given
in~(\ref{shockmetric}). By virtue of the
identities~(\ref{invvanish}) the operator ${\rm H}_{\mu\nu}$
vanishes, ensuring that the square curvature scalar does not
contribute to the impulsive wave solution. The field
equations~(\ref{fieldeq}), reduces to the following linear
fourth-order two dimensional partial differential equation:
\begin{eqnarray}
-\frac{1}{2}[\beta\nabla^4_\perp+\nabla^2_\perp]H(x^i,u)=8\pi
GT_{uu},\label{feqH}
\end{eqnarray}
where $\nabla^4_\perp=\nabla^2_\perp\nabla^2_\perp$. In the
Einstein gravity, $\alpha=\beta=0$, the equation~(\ref{fieldeq})
reduces to
\begin{equation}
-\frac{1}{2}\nabla^2_{\perp}H_{\rm E}(x^i,u)=8\pi
GT_{uu}(x^i,u).\label{einsppeq}
\end{equation}
Taking into account the above equation we can rewrite~(\ref{feqH}) as
\begin{equation}
[\beta\nabla^2_\perp+1]H(x^i,u)=H_{{\rm E}}(x^i,u)+ah(x^i,u),
\end{equation}
where $h(x^i,u)$ is a harmonic function of the transverse coordinates $x^i$ and, 
$a$ an arbitrary constant. This result means that we can add to the final 
solution a harmonic function of the transverse coordinates. The particular 
dependence of $h$ on $x^i$ is dictated by the symmetries of the problem and the 
dependence on the $u$ coordinate is determined by the source term. 
By making a decomposition of $H(x^i,u)$ as
\begin{equation}
H(x^i,u)= H_{\rm Q}(x^i,u)+H_{{\rm E}}(x^i,u)+ah(x^i,u),
\end{equation}
where the index ${\rm Q}$ refers to the purely quadratic part of the solution,
we obtain the following second order partial differential equation for $H_{\rm
Q}(x^i,u)$:
\begin{equation}
\left[\nabla^2_\perp+\frac{1}{\beta}\right]H_{Q}(x^i,u)=16\pi
GT_{uu}.\label{QeqH}
\end{equation}
Thus, the problem of solve the fourth order equation~(\ref{feqH}) was reduced to 
the problem of solve the second order equations~(\ref{einsppeq}) 
and~(\ref{QeqH}).

\section{The gravitational shock wave associated with a homogeneous beam of
null particles} \label{sec-3}

In this section, we compute the solution of
equations~(\ref{einsppeq}) and~(\ref{QeqH}) when the source is
given by a distribution of null particles moving along the same
direction. The relevant component of the corresponding energy
momentum tensor in polar coordinates is given by~\cite{baris1}:
\begin{eqnarray}
T_{uu}=\lambda\varrho(r,\theta)\delta(u),\label{Tuu}
\end{eqnarray}
where $\varrho(r,\theta)$ is the energy density of the distribution of
null particles and $\lambda$ is a dimensionless constant which varies within the 
range $1\leq\lambda\leq 2$ and comes from the equation of state
\begin{equation}
P=(\lambda-1)\varrho,
\end{equation}
which describes an ultrarelativistic perfect fluid~\cite{chop}.

The line element~(\ref{shockmetric}) can be written as
\begin{eqnarray}
ds^2=-dudv+[f_{\rm E}(r,\theta)+f_{\rm
Q}(r,\theta)]\delta(u)du^2+dr^2+r^2d\theta^2, \label{linerteta}
\end{eqnarray}
where the functions $f_{\rm E}$ and $f_{\rm Q}$ must satisfy the
following equations:
\begin{eqnarray}
\nabla^2f_{\rm E}(r,\theta)=-16\pi G\lambda\varrho(r,\theta)
\label{feeq}
\end{eqnarray}
and
\begin{eqnarray}
\left[\nabla^2+\frac{1}{\beta}\right]f_{\rm Q}(r,\theta)=16\pi
G\lambda\varrho(r,\theta).\label{fQeq}
\end{eqnarray}

We assume that the energy density is a constant $\varrho_0$ 
within a certain radius $0\leq r\leq R_0$ and zero outside. Thus, the source 
represents a cylindrical beam of null particles with width $R_0$.
This beam  is a simple generalization of a single null particle. 
We know that the wave profile of a gravitational shock wave generated by a 
single null particle in quadratic gravity is given by
\begin{equation}
f_0(r)=8 G p 
\left[\ln\left(\frac{r}{r_0}\right)+K_0\left(\frac{r}{\sqrt{-\beta}}\right)
\right],
\label{spsol}
\end{equation}
where $p$ is  momentum of the particle, $K_0$ a modified Bessel function and 
$r_0$ some integration constant~\cite{lou1}. This is a continuos function of $r$ which is regular at the origin an diverges logarithmically at $r\rightarrow\infty$. 

To obtain a solution which generalizes the gravitational shock wave generated by 
a single null particle we proceed by solving the equations~(\ref{feeq}) 
and~(\ref{fQeq}) choosing the integration constants in such a way that the 
wave profile $f(r)=f_{\rm E}(r)+f_{\rm Q}(r)$ fulfill the same boundary and 
regularity conditions that the single null particle shock wave 
profile~(\ref{spsol}). Therefore, $f(r)$ must be {\it i)} a continuous function of 
$r$ (~even at $r=R_0$, where the source have a discontinuity~), {\it ii)} 
regular at the origin, namely $f(0)<\infty$, and {\it iii)} logarithmically 
divergent at $r\rightarrow\infty$. 
 
Let us solve the equation~(\ref{feeq}). 
The
cylindrical symmetry implies that $f_{\rm E}$ will
depend only on the $r$ coordinate. It is easy to see that the solution 
to~(\ref{feeq}) can be written as
\begin{equation}
f_{\rm E}(r)=\cases{-4\pi G\lambda\varrho_0r^2+C_1\ln(r)+C_2\qquad 
&for\qquad$r\leq R_0$\\
D_1\ln(r)+D_2\qquad 
&for$\qquad r>R_0$.}
\label{fesolc}
\end{equation}
where $C_1$, $C_2$, $D_1$ and $D_2$ are integration constants. 

By intergrating both sides of~(\ref{feeq}), with $f_{\rm E}(r)$ given 
by~(\ref{fesolc}), over a circle of radius $r<R_0$ and applying the divergence 
theorem to the left-hand side term we get $C_1=0$. Now, integrating over a 
circle of radius $r>R_0$ and applying the divergence theorem to the left-hand 
side term
we obtain $D_1=-8\pi G\lambda\varrho_0R_0^2$. The remaining 
integration constants $C_2$ and $D_2$, are determined by the boundary and 
regularity conditions. As a regularity condition we impose the continuity of 
$f_{\rm E}(r)$ at $r=R_0$.
This condition imply that
\begin{eqnarray}
D_2=16\pi G\lambda\varrho_0 
R_0^2\left(\frac{1}{2}\ln(R_0)-\frac{1}{4}\right)+C_2.\nonumber
\end{eqnarray}
The constant $C_2$, which gives the value of $f_{\rm E}$ at $r=0$, is the only 
arbitrary constant that will be determined by the boundary conditions. 
The regularity at the origin is satisfied by given any finite value to $C_2$, Then, for simplicity and without loose of 
generality we choose $C_2=0$. Thus, the solution of equation~(\ref{feeq}) for 
any $r\geq 0$ is given by

\begin{equation}
f_{\rm E}(r)=\cases{-4\pi G\lambda\varrho_0r^2\qquad &for\qquad$r\leq R_0$\\
-8\pi
G\lambda\varrho_0R_0^2\left[\ln\left(\frac{r}{R_0}\right)+
\frac{1}{2}\right]\qquad &for$\qquad r>R_0$.}
\label{fesol}
\end{equation}

Let us turn out to the purely quadratic gravity part of the wave profile,  
$f_{\rm
Q}(r)$. Integration of the equation~(\ref{fQeq})
leads to the following result~\cite{grad}:
\begin{equation}
f_{\rm
Q}(r)\!=\!\cases{A_1 
K_0\left(\frac{r}{b}\right)+A_2I_0\left(\frac{r}{b}\right)-16\pi
G\lambda\varrho_0b^2\qquad &for$\qquad r\leq R_0$\\
\\
B_1K_0\left(\frac{r}{b}\right)+B_2I_0\left(\frac{r}{b}\right)\qquad &for$\qquad 
r>R_0$,}
\label{fqsolc}
\end{equation}
where $I_{\nu}$ and $K_{\nu}$ are modified Bessel functions,
$b\equiv\sqrt{-\beta}$ and $A_1$, $A_2$, $B_1$ and $B_2$ are integration 
constants.

To obtain a solution $f(r)$ regular at the origin and logarithmically divergent at $r\rightarrow\infty$, we set $B_2=0$ and $A_1=0$ in~(\ref{fqsolc}). As we want that $f(r)$ be a 
continuous function of $r$, $f_{\rm Q}(r)$ must be continuous at $r=R_0$. This condition is fulfilled by taking
\begin{eqnarray}
A_2=16\pi G\lambda\varrho_0bR_0K_1\left(\frac{R_0}{b}\right)\nonumber
\end{eqnarray} 
and
\begin{eqnarray}
B_1=-16\pi G\lambda\varrho_0bR_0I_1\left(\frac{R_0}{b}\right). \nonumber
\end{eqnarray}
Thus, $f_{\rm Q}(r)$ takes the following form    
\begin{equation}
f_{\rm
Q}(r)=\cases{16\pi
G\lambda\varrho_0\left[R_0{b}K_1\!\left(\frac{R_0}{b}\right)\!I_0\!\left(\frac{r
}{b}\right)\!-\!b^2\right]\qquad &for$\qquad r\leq R_0$\\
\\
-16\pi
G\lambda\varrho_0R_0{b}I_1\!\left(\frac{R_0}{b}\right)\!K_0\!\left
(\frac{r}{b}\right)\qquad &for$\qquad r> R_0$,}
\label{foutqsol}
\end{equation}

The solution to equation~(\ref{feqH}) with the source term~(\ref{Tuu}) 
is given by
\begin{equation}
H(r,u)=[f(r)+a { h}(r)]\delta(u),
\label{solh}
\end{equation}
where $f(r)=f_{\rm E}(r)+f_{\rm Q}(r)$ is the wave profile and ${ h}(r)$ is a harmonic function.
The cylindrical symmetry implies that ${ h}(r)=\ln(r)$. Note that $h(r)$ diverges at $r=0$ and $r=\infty$. If we choose $a=-8\pi G\lambda\varrho_0R_0^2$ 
the wave profile approaches to a constant value of $a/2$ as 
$r\rightarrow\infty$. However, the solution will diverge at $r=0$ unless that we 
put $A_1=-a$. But, this choice violate the previous choice, $A_1=0$, and implies in the lack of continuity of $f_{\rm Q}(r)$ at 
$r=R_0$. Therefore, to fulfill the boundary and regularity conditions chosen for $f(r)$ we must set $a=0$ in~(\ref{solh}). The final result 
diverges logarithmically at $r\rightarrow\infty$ and resembles the wave profile 
for a gravitational shock wave generated by a single null particle~\cite{lou1}

\begin{figure}
\begin{center}
\epsfig{figure=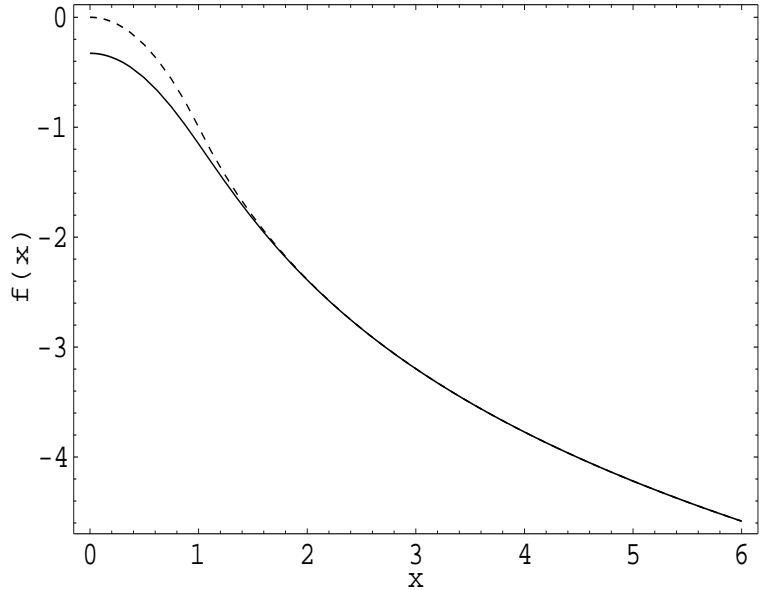,angle=360,height=8cm,width=10cm}
\caption{\label{figur1}A comparison between the wave profiles in
Einstein's gravity (dashed line) and for quadratic gravity (solid
line). To clarify the difference of the predictions given by
Einstein and quadratic gravity, we set $4\pi G\lambda\varrho_0=1$,
$R_0=1$ and plot the  $f(r)$ curve for $\sqrt{-\beta}=0.3$. We
note a non zero residual value at $r=0$ for the quadratic theory
and, we als note that the two curves becomes indistinguishable
after a few $R_0$ distance.}
\end{center}
\end{figure}

The figure~\ref{figur1}, shows the $r$ dependence of the wave
profiles $f_{\rm E}(r)$ (dashed line) and $f(r)=f_{\rm E}(r)+f_{\rm Q}(r)$ 
(solid line).

\noindent
The non zero value of
$f(r)$ at $r=0$ is given by
\begin{equation}
f(0)=16\pi G\lambda\varrho_0\left[{b} R_0
K_1\!\left(\frac{R_0}{{b}}\right)-b^2\right].
\end{equation}
This quantity results from the Ricci-squared term in the gravitational 
Lagrangian.
It carries the information of the beam width $R_0$ and tends to zero as
$R_0\rightarrow 0$.
We also
observe that the wave profile obtained for quadratic gravity agrees
with the prediction of Einstein's gravity at large distances from
the beam axis.
\begin{figure}
\begin{center}
\epsfbox{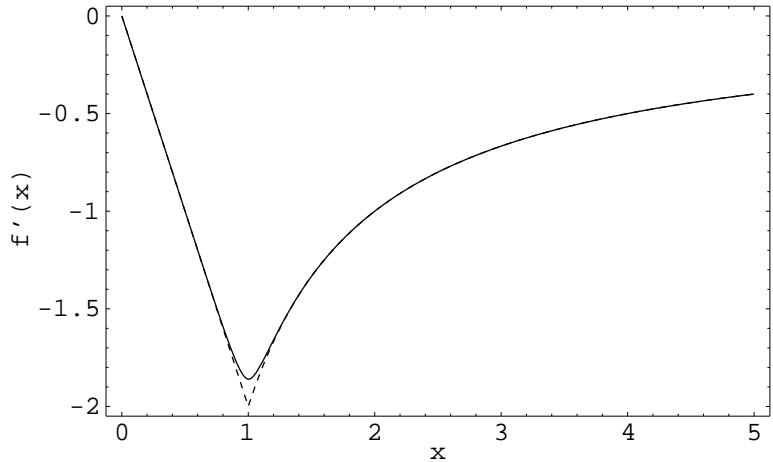} \caption{\label{figur2} In this figure we plot
$f^\prime(r)$ (solid line) and $f_{\rm E}^\prime(r)$ (dashed line)
as function of the polar coordinate $r$. The units are the same
used to plot figure~\ref{figur1}, but here we set
$\sqrt{-\beta}=0.07$ to compute $f^\prime(r)$. We note the
softness of the quadratic gravity result at $r=R_0$ in comparison
to the Einstein's gravity one.}
\end{center}
\end{figure}

\begin{figure}
\begin{center}
\epsfbox{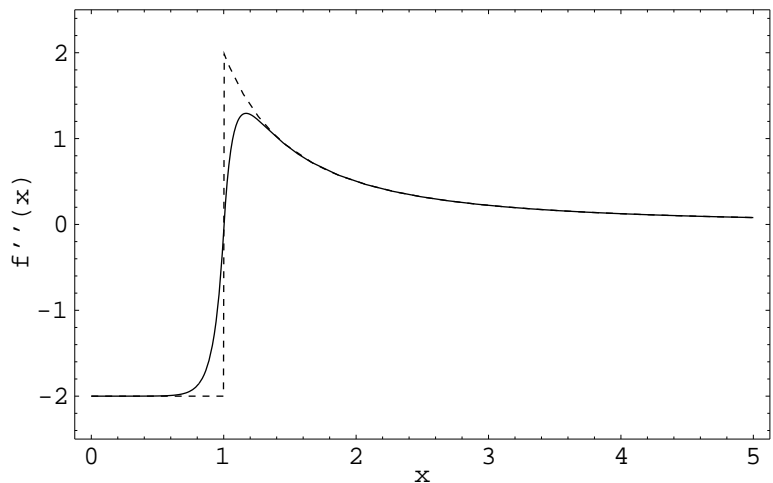} \caption{\label{figur3} Here are depicted the
$f^{\prime\prime}(r)$ (solid line) and $f_{\rm
E}^{\prime\prime}(r)$ (dashed line) as function of $r$ using the
units of figure~\ref{figur1} with $\sqrt{-\beta}=0.07$. We note
that the quadratic gravity curve is smooth at $r=R_0$ while the
Einstein's gravity curve shows a discontinuity at this point.}
\end{center}
\end{figure}

A comparison between the first and second derivatives of the wave
profiles obtained in the two theories shows that the higher
curvature terms of quadratic gravity contributes to smooth out the
$r$ dependence of these functions at $r=R_0$. This ``smoothing'' property of 
quadratic gravity is related to the fact that the field equations are of fourth 
order in quadratic gravity, whereas they are of second order in Einstein's 
Gravity.

In figure~\ref{figur2} we plot the first derivative with respect
to $r$ of the wave profiles in both theories. The dashed line
represents $f_{\rm E}^\prime(r)$ and the solid line represents
$f^\prime(r)$ (the {\it prime} denotes the derivative with respect
to $r$). Figure~\ref{figur3} shows the $r$ dependence of the
second derivatives of the wave profiles. We note the continuity
and softness of $f^{\prime\prime}(r)$ at the point in which
$f^{\prime\prime}_{\rm E}(r)$ has a discontinuity. Therefore, we
can conclude that the quadratic curvature component associated to
the Ricci-squared term in the gravitational action smooths out the
functional dependence of the derivatives of the wave profile.
Furthermore, this higher-curvature term contribute to remove the
discontinuity of the second derivative of the Einstein's gravity
wave profile at $r=R_0$.

We must bear in mind that a gravitational shock wave cannot be treated in the 
same way as a linearized gravitational wave. The effect of a gravitational shock 
wave on test particles is a subject that we will treat in an incoming paper. 
Nevertheless, we can remark that the test particles would be be sensitive to the 
analytic proprieties of the wave profile function~\cite{stein}. Obviously, these 
properties remains the same whatever the particular values assumed for the 
$\beta$ parameter. Thus, even at the limit of $\sqrt{-\beta}\rightarrow 0$ the 
quadratic gravity could in principle  be distinguished from Einstein's gravity.

\section{Final remarks}
\label{conc} Let us now summarize the main results obtained in our
investigation and give a suggestion for application of these results in 
astrophysics.

The first remarkable result is that the $R^2$
quadratic invariant does not contribute to the wave solution. The
only contribution, which comes from quadratic part of the theory,
comes from the Ricci-squared term in the quadratic gravity action.
The effect of this term over the gravitational wave emission is to
regularize the discontinuity associated to the Einstein's gravity
solution.
We recall that at large distances from the beam axis the predictions 
derived in  both theories coincide.

The gravitational shock wave is very different as compared to an ordinary gravitational wave~\cite{barhog1}. The geodesics and the geodesic deviations in spacetimes of gravitational shock waves have been studied in~\cite{stein}. The authors have showed that the behavior of small geodesic deviations will be dependent on the combinations of the wave profile and theirs derivatives in the transverse space with products and powers of the Dirac $\delta$ distribution and the ``kink'' function, defined by $u\theta(u)$, where $\theta(u)$ is the step function. Therefore, the effect of a gravitational shock wave on test particles will be dependent on these results. We consider that the issue of observation of gravitational shock waves deserves a separated investigation and we intend to do so in another paper, now in preparation, to appear elsewhere. 

An important question remains to be discussed. Can the model studied in present the work be addressed to the study of astrophysical sources?  We know that some GRBs are generated in the expansion of ultrarelativistic jets~\cite{saripiran99,usov94}. Then a cylindrical beam of null particles can be used as a frist approximation to an ultrarelativistic jet. In this approximation 
an exact shock wave solution was 
obtained. We compute the wave forms in the framework of Einstein's and quadratic gravity. If a gravitational shock wave takes place in a GRB 
scenario it may have important consequences to the physical process at the source. According to~\cite{garver1} a gravitational 
shock wave would be relevant to the  high energy scattering of quantum particles. Therefore it can be espected that the gravitational shock wave  contributes to the dynamics of a GRB source and to the composition of the ejected material. Furthermore, once a gravitational shock wave related to a 
GRB can be detected, the results derived in the present work could contribute to the understanding of the ultrarelativistic jets features from their gravitational shock wave profiles.

\ack{We thank Dr. J.A. de Freitas Pacheco for stimulating
discussions. We also thank the Brazilian agency FAPESP (grants 00/10374-5, 
97/13720-7, 97/06024-4)
and CNPq (grants 300619/92-8, 380.503/02-6) for the
financial support.}

\end{document}